\documentclass[%
 reprint,
 amsmath,amssymb,
 aps,
]{revtex4-2}

\usepackage{pkgfile}
\usepackage{xifthen}
\usepackage{tabularx}
\usepackage{adjustbox}
\usepackage{bbm}
\usepackage{makecell}

\usepackage{tikz}
\usetikzlibrary{matrix,arrows}

\usepackage{graphicx}
\usepackage{dcolumn}
\usepackage{bm}

\begin{document}


	\title[Boundary Condition and Phase Factors in Feynman Path Integral] {Boundary Condition and the Auxiliary Phase in Feynman Path Integral}

\author{Chung-Ru Lee}
\affiliation{Department of Mathematics\\
National University of Singapore.}
\email{crlee@nus.edu.sg}

\date{\today}
\counterwithout{equation}{section}

\begin{abstract}
When employing Feynman path integrals to compute propagators in quantum physics, the concept of summing over the set of all paths is not always na\"ive. In fact, an auxiliary phase often has to be included as a weight for each summand. In this article we discuss the nature of those phase factors for the various types of boundary conditions including all three of the Dirichlet, Neumann and Robin types, as well as their mixtures. We verify that for a free particle confined on a line segment, the resulting formula on the propagator matches those arising from the Schr\"{o}dinger eqaution, with a trivial normalization factor.
\end{abstract}
\date{\today}
\maketitle


\section{Introduction}
The wave function of a free particle trapped in an infinite potential well has been studied thoroughly using both the Schr\"odinger equation and Feynman path integral \cite{Goodman,Gasiorowicz,Griffiths}.

Fix some $L>0$ and let $$H=\frac{p^2}{2m}+V(x)=-\frac{\hbar^2}{2m}\frac{d^2}{dx^2}+V(x),$$
where the potential $V(x)=\begin{cases}0 & 0\leq x\leq L\\
\infty & \text{otherwise}\end{cases}$.
The corresponding (time-independent) Schr\"odinger equation is the eigenvalue problem
\begin{align}
H\psi=E\psi\label{eqn:eigenvalue_problem}
\end{align}
subject to the Dirichlet boundary conditions $\psi(0)=\psi(L)=0$.

On the spectral end, the arising eigenvalues $E$ are discrete. One simply label them as $E_n$ (each with multiplicity one) and the corresponding eigenfunctions as $\psi_n=\left|n\right>$. In fact, after normalization
$$\psi_n(x)=\sqrt{\frac{2}{L}}\sin(k_nx),$$
where $k_n=\frac{n\pi}{L}$ and therefore $E_n=\frac{\hbar^2}{2m}k_n^2=\frac{n^2\pi^2\hbar^2}{2mL^2}$ for $n=1,2,3,\dots$.

Decomposed by the spectrum of $H$, the propagator can then be computed by
\begin{align}
K(y,t_1;x,t_0)=\frac{2}{L}\sum_{n=1}^\infty e^{\frac{i}{\hbar}E_n(t_1-t_0)}\sin(k_nx)\sin(k_ny).\label{eqn:dirichlet_schrodinger}
\end{align}

On the other hand, since this is a case with a quadratic ambient potential, it suffices to consider only the contribution from all classical paths \cite{Feynman,MacKenzie,Schulman,Kleinert,zinn-justin}. The expansion along the momentum $p$ gives an expression for $K(y,t_1;x,t_0)$:
\begin{align*}
\sum_{r=-\infty}^\infty\frac{\epsilon_r}{2\pi\hbar}\int_{\mathbb{R}} e^{-\frac{i}{\hbar}\frac{p^2}{2m}(t_1-t_0)}e^{\frac{i}{\hbar}p(y_r-x)}dp.
\end{align*}
where the sum over $r$ is the sum indexed by possible classical paths (see Figure 1), and $\epsilon_r$ is a unitary phase factor associated to the path to $y_r$. Detail of these derivation can be found in \cite{Goodman,Janke-Kleinert,Soekmen}, where they pointed out that the $\epsilon_r$ above should be $(-1)^r$. Under such principle, the results derived from the two ways of computing the propagator match.

One disparity between the two approaches lies within the assignment of phases. Since Schr\"odinger equation deals with wave functions, it is by construction that phases are included when considering the interaction between states. On the other hand, even though path integral taken all paths into account, so far it does not seem to exist an intrinsic way of associating phases to each path in general. Related discussion can be found in \cite{Dowker,Mouchet,Suzuki}.

To further study this auxiliary phase factor, certain modification on the model has to be made. Instead of considering the infinite potential well, one consider the wave function of an $1$-dimensional free particle confined to a line segment $[0,L]$. This interpretation makes it sensible to discuss other boundary conditions such as the Neumann boundary conditions and so on.

In particular, the original problem of an infinite potential well can be regarded as the special case of imposing the Dirichlet boundary conditions on both ends. This article aims to state and apply the Principle of Reflection, which we will summarize in the next section.

\section{A Guideline}\label{sec:principle}
From a classical point of view, this phase change should be a local factor and is dependent only upon the condition around the boundary point.

\begin{prin*}
The phase factor associated to each reflection can be determined using either the model of a free particle on a ray with suitable boundary conditions imposed (details in the following passage). The auxiliary phase that occurs as the weight for each classical path in the sum equals the product over the local phase factor for all reflections of that path, relative to the initial direction of that path.
\end{prin*}

The statement \textit{relative to the initial direction} means that the product over the local phase factors should be projected onto the initial normal direction (in 1-dimensional models, it is just a multiplication of $\pm 1$ on the angle $\theta$).

The Principle above can be seen as a generalization to the method of image point formulation described in \cite{Goodman}, which described the picture when $\epsilon_k=-1$ for all reflections. If every $\epsilon_k$ from above are $\pm 1$, just as in the previous cases, then the factor is equal to the simple product of local factors.

Therefore, to further examine the Principle, it is necessary to consider boundary conditions that would cause a non-trivial phase to arise.

\section{Neumann Boundary Conditions}\label{sec:neumann}
In this section we consider (\ref{eqn:eigenvalue_problem}) with the boundary conditions $\psi'(0)=\psi'(L)=0$. Once again, the eigenvalues in this case are discrete multiplicity-free and will be labelled as $E_n$. The corresponding eigenfunctions are
$$\psi_n(x)=\sqrt{\frac{2}{L}}\cos(k_nx)$$
with $k_n=\frac{n\pi}{L}$, $E_n=\frac{n^2\pi^2\hbar^2}{2mL^2}$ for $n=1,2,\dots$. Note that here the case $n=0$ also contributes a nontrivial solution $\psi_0=\sqrt{\frac{1}{L}}$.

Decompose by the spectrum of $H$, the propagator can $K(y,t_1;x,t_0)$ thereby be computed by
\begin{align*}
\frac{1}{L}+\frac{2}{L}\sum_{n=1}^\infty e^{-\frac{i}{\hbar}E_n(t_1-t_0)}\cos(k_nx)\cos(k_ny).
\end{align*}

Performing the path integral calculation gives another expression for $K(y,t_1;x,t_0)$:
\begin{align}
\frac{1}{2\pi\hbar}\sum_{r=-\infty}^\infty\epsilon_r\int_{\mathbb{R}} e^{-\frac{i}{\hbar}\frac{p^2}{2m}(t_1-t_0)}e^{\frac{i}{\hbar}p(y_r-x)}dp.\label{eqn:path_integral_elementary}
\end{align}

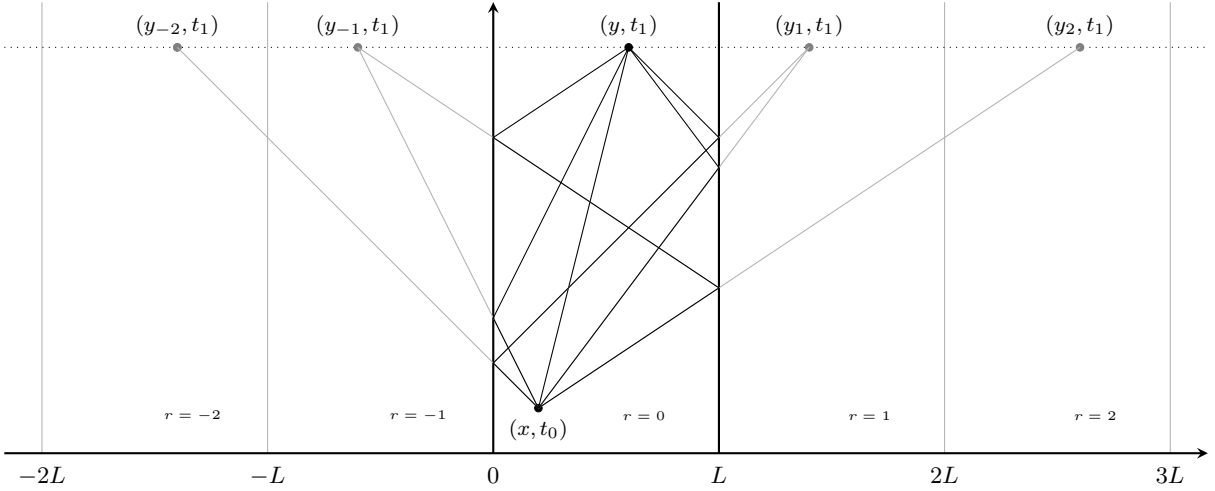
\begin{figure*}[ht]
\centering
\begin{tikzpicture}
\foreach \x in {-2,-1,...,3}
\draw [thin, black!30] (3*\x,0) -- (3*\x,6);
\foreach \x in {-2,-1,...,2}
\node at (3*\x+2,0.5) {\tiny $r=\x$};
\node[below=2pt] at (0,0) {$0$};
\node[below=2pt] at (3,0) {$L$};
\node[below=2pt] at (6,0) {$2L$};
\node[below=2pt] at (9,0) {$3L$};
\node[below=2pt] at (-3,0) {$-L$};
\node[below=2pt] at (-6,0) {$-2L$};
\draw [-stealth,thick] (-6.5,0) -- (9.5,0);
\draw [-stealth,thick] (0,0) -- (0,6);
\draw [black,thick] (3,0) -- (3,6);
\node[draw,circle,inner sep=1pt,fill,black] at (0.6,0.6) {};
\node[below=1pt] at (0.6,0.6) {\footnotesize $(x,t_0)$};
\node[draw,circle,inner sep=1pt,fill,black] at (1.8,5.4) {};
\node[above=1pt] at (1.8,5.4) {\footnotesize$(y,t_1)$};
\draw [dotted] (-6.5,5.4) -- (9.5,5.4);
\node[draw,circle,inner sep=1pt,fill,black!50] at (4.2,5.4) {};
\node[above=1pt] at (4.2,5.4) {\footnotesize$(y_1,t_1)$};
\node[draw,circle,inner sep=1pt,fill,black!50] at (7.8,5.4) {};
\node[above=1pt] at (7.8,5.4) {\footnotesize$(y_2,t_1)$};
\node[draw,circle,inner sep=1pt,fill,black!50] at (-1.8,5.4) {};
\node[above=1pt] at (-1.8,5.4) {\footnotesize$(y_{-1},t_1)$};
\node[draw,circle,inner sep=1pt,fill,black!50] at (-4.2,5.4) {};
\node[above=1pt] at (-4.2,5.4) {\footnotesize$(y_{-2},t_1)$};
\node[draw,circle,inner sep=1pt,fill,black!50] at (-4.2,5.4) {};
\draw[black] (0.6,0.6) -- (1.8,5.4);
\draw[thin,black!30] (0.6,0.6) -- (4.2,5.4);
\draw[thin,black!30] (0.6,0.6) -- (7.8,5.4);
\draw[thin,black!30] (0.6,0.6) -- (-1.8,5.4);
\draw[thin,black!30] (0.6,0.6) -- (-4.2,5.4);
\draw[thin,black!30] (3,2.2) -- (-1.8,5.4);
\draw[thin,black!30] (0,0.6+4.8/8) -- (4.2,5.4);
\draw[black] (0.6,0.6) -- (3,3.8) -- (1.8,5.4);
\draw[black] (0.6,0.6) -- (3,2.2) -- (0,4.2) -- (1.8,5.4);
\draw[black] (0.6,0.6) -- (0,1.8) -- (1.8,5.4);
\draw[black] (0.6,0.6) -- (0,0.6+4.8/8) -- (3,0.6+4.8*3/4) -- (1.8,5.4);
\end{tikzpicture}
\caption{Several of the classical paths connecting $(x,t_0)$ and $(y,t_1)$}
\end{figure*}

The Reflection Principle indicates that $\epsilon_r=1$ in this case. Notice that
$$y_r=\begin{cases}
rL+y & \text{for even }r\\
(r+1)L-y & \text{for odd }r.
\end{cases}$$
Reset the parameter modulo $2$ as $r=2l$ for even $r$ or $2l-1$ for odd $r$ with $l\in\mathbb{Z}$ and plugging into (\ref{eqn:path_integral_elementary}), $K(y,t_1;x,t_0)$ is then
\begin{align*}
\frac{1}{\pi\hbar}\int_{-\infty}^\infty\sum_{l=-\infty}^\infty e^{\frac{2ilLp}{\hbar}}e^{-\frac{i}{\hbar}\frac{p^2}{2m}(t_1-t_0)}e^{-\frac{i}{\hbar}px}\cos(py/\hbar)dp
\end{align*}
By Poisson summation formula,
\begin{align*}
\int_{-\infty}^\infty\sum_{l=-\infty}^\infty & e^{2ilLp/\hbar}f(p)dp\\
&=\frac{\pi\hbar}{L}\sum_{n=-\infty}^\infty f(\frac{n\pi\hbar}{L})=\frac{\pi\hbar}{L}\sum_{n=-\infty}^\infty f(k_n\hbar).
\end{align*}
Thus $K(y,t_1;x,t_0)$ equals
\begin{align}
& \frac{1}{L}\sum_{n=-\infty}^\infty e^{-\frac{i}{\hbar}E_n(t_1-t_0)}e^{-ik_nx}\cos(k_ny)\notag\\
= & \frac{2}{L}\left(\frac{1}{2}+\sum_{n=1}^\infty e^{-\frac{i}{\hbar}E_n(t_1-t_0)}\cos(k_nx)\cos(k_ny)\right).\label{eqn:poisson}
\end{align}
The last equation (\ref{eqn:poisson}) follows from writing $e^{-ik_nx}=\cos(k_nx)-i\sin(k_nx)$ and observe the parity of the functions. The formula above matches the result of the spectral decomposition.

\section{Robin Boundary Conditions}\label{sec:robin}

Following the previous examples, we consider (\ref{eqn:eigenvalue_problem}) with the Robin boundary conditions $\psi(0)+\alpha\psi'(0)=\psi(L)+\alpha\psi'(L)=0$ (be cautious of the signs chosen here). The eigenfunctions are labelled as $\psi_n$. So the wave function $\psi_n(x)$ equals
$$(\frac{L}{2})^{-1/2}\frac{1}{\sqrt{1+\alpha^2k_n^2}}\big(-\alpha k_n\cos(k_nx)+\sin(k_nx)\big)$$
with $k_n=n\frac{\pi}{L}$ and $E_n=n^2\frac{\pi^2\hbar^2}{2mL^2}$ for $n=1,2,\dots$.

By the spectral decomposition,
$$K(y,t_1;x,t_0)=\sum_{n=1}^\infty e^{-\frac{i}{\hbar}E_n(t_1-t_0)}\overline{\psi_n(x)}{\psi_n(y)}.$$
The path integral approach derived the propagator from the (weighted) sum over all classical paths
$$\frac{1}{2\pi\hbar}\sum_{r=-\infty}^\infty\int_{\mathbb{R}} \epsilon_re^{-\frac{i}{\hbar}\frac{p^2}{2m}(t_1-t_0)}e^{\frac{i}{\hbar}p(y_r-x)}dp.$$
Notice that
$$y_r=\begin{cases}
rL+y & \text{for even }r\\
(r+1)L-y & \text{for odd }r.
\end{cases}$$
With the Reflection Principle in \S\ref{sec:principle}, since the paths parameterized by even $r$ pass through (in the sense of the image point method) an even number of alternating boundaries, the cancellation yields $\epsilon_r=1$. Meanwhile, a single phase factor $\epsilon_r=e^{i\theta}=-\frac{1-i\alpha k}{1+i\alpha k}$ remains after cancellation when $r$ is odd. Note that $p=\hbar k$, so the auxiliary phase $\epsilon$ depends on the momentum of the classical path here.

We may therefore rewrite the propagator as
\begin{widetext}
\begin{align*}
K(y,t_1;x,t_0) & =\frac{1}{2\pi\hbar}\sum_{r=-\infty}^\infty\int_{-\infty}^\infty\left(e^{-\frac{i}{\hbar}\frac{p^2}{2m}(t_1-t_0)}e^{\frac{i}{\hbar}p(2rL+y-x)}+e^{i\theta}e^{-\frac{i}{\hbar}\frac{p^2}{2m}(t_1-t_0)}e^{\frac{i}{\hbar}p(2rL-y-x)}\right)dp\\
&=\frac{1}{2\pi\hbar}\int_{-\infty}^\infty\sum_{l=-\infty}^\infty e^{2ilLp/\hbar}e^{-\frac{i}{\hbar}\frac{p^2}{2m}(t_1-t_0)}e^{-\frac{i}{\hbar}px}(e^{-\frac{i}{\hbar}py}+e^{i\theta}e^{\frac{i}{\hbar}py})dp.
\end{align*}
\end{widetext}

Employing the Poisson summation formula used in \S\ref{sec:neumann}, we obtain the energy levels $E_n=\frac{p_n^2}{2m}=\frac{\hbar^2}{2m}k_n^2$, where $k_n=\frac{n\pi}{L}$. After simplification (see Appendix for detailed computations), we find that $K(y,t_1;x,t_0)$ is propotional to
\begin{align}
&\sum_{n=-\infty}^\infty e^{-\frac{i}{\hbar}E_n^2(t_1-t_0)}e^{-ik_nx}\left(e^{ik_ny}+e^{i\theta}e^{-ik_ny}\right)\notag\\
= & \sum_{n=-\infty}^\infty e^{-\frac{i}{\hbar}E_n(t_1-t_0)}\cos(k_nx-\frac{\theta}{2})\cos(k_ny-\frac{\theta}{2})\label{eq:robin}.
\end{align}
Furthermore,
$$\cos(k_nx-\frac{\theta}{2})=\cos(k_nx)\cos(\frac{\theta}{2})+\sin(k_nx)\sin(\frac{\theta}{2})$$
while
$$\cos(\frac{\theta}{2})/\sin(\frac{\theta}{2})=-\alpha k_n.$$
Substitute in the previous equation (\ref{eq:robin}) and normalize. The result agrees with the result from spectral decomposition.

\section{Mixed Boundary Conditions}

To further examine whether the concept of the phase factor is robust,
we now consider (\ref{eqn:eigenvalue_problem}) with the boundary conditions $\psi'(0)=\psi(L)=0$. That is, Neumann on one side and Dirichlet on the other.

In this case, the eigenfunctions, again labelled as $\psi_n$ are given by
$$\psi_n(x)=\sqrt{\frac{2}{L}}\cos(k_nx)$$
with $k_n=(n-\frac{1}{2})\frac{\pi}{L}$ and $E_n=(n-\frac{1}{2})^2\frac{\pi^2\hbar^2}{2mL^2}$ for $n=1,2,\dots$.
The spectral decomposition tells us that $K(y,t_1;x,t_0)$ equals
$$\frac{2}{L}\sum_{n=1}^\infty e^{-\frac{i}{\hbar}E_n(t_1-t_0)}\cos(k_nx)\cos(k_ny).$$
The path integral approach considers the (weighted) sum over classical paths
$$\frac{1}{2\pi\hbar}\sum_{r=-\infty}^\infty\epsilon_r\int_{\mathbb{R}} e^{-\frac{i}{\hbar}\frac{p^2}{2m}(t_1-t_0)}e^{\frac{i}{\hbar}p(y_r-x)}dp.$$
The Reflection Principle indicates that $\epsilon_r=1$ if $r=-1,0$ modulo $4$ and $\epsilon_r=-1$ if $r=1,2$ modulo $4$. We can therefore rewrite the sum as
\begin{align*}
\frac{1}{\pi\hbar}\int_{-\infty}^\infty\sum_{l=-\infty}^\infty & e^{4ilLp/\hbar}e^{-\frac{i}{\hbar}\frac{p^2}{2m}(t_1-t_0)}e^{-\frac{i}{\hbar}px}\\
&\cos(py/\hbar)(1-e^{2iLp/\hbar})dp.
\end{align*}

Employing the Poisson summation formula,
$$\int_{-\infty}^\infty\sum_{l=-\infty}^\infty e^{4ilLp/\hbar}f(p)dp=\frac{\pi\hbar}{2L}\sum_{b=-\infty}^\infty f(\frac{b\pi\hbar}{2L}).$$
Thus $K(y,t_1;x,t_0)$ equals
\begin{align*}
\frac{1}{2L}\sum_{b=-\infty}^\infty & e^{-\frac{i}{\hbar}\frac{1}{2m}(\frac{b\pi\hbar}{2L})^2(t_1-t_0)}\\
& e^{-ib\pi x/2L}\cos(b\pi y/2L)(1-e^{ib\pi}).
\end{align*}
Note that $e^{ib\pi}=(-1)^b$, so $$1-e^{ib\pi}=\begin{cases}2 & \text{for odd }b\\
0 & \text{for even }b\end{cases}.$$
Writing the odd $b$ as $2n-1$, we have $\frac{1}{2m}(\frac{b\pi\hbar}{2L})^2=\frac{\pi^2\hbar^2}{2mL^2}(n-\frac{1}{2})^2=E_n$ and $\frac{b\pi}{2L}=\frac{\pi}{L}(n-\frac{1}{2})=k_n$. That is, $K(y,t_1;x,t_0)$ equals
\begin{align}
& \frac{1}{L}\sum_{n=-\infty}^\infty e^{-\frac{i}{\hbar}E_n(t_1-t_0)}e^{-ik_nx}\cos(k_ny)\notag\\
=& \frac{2}{L}\sum_{n=-\infty}^\infty e^{-\frac{i}{\hbar}E_n(t_1-t_0)}\cos(k_nx)\cos(k_ny).
\end{align}
Again, this matches the result from spectral decomposition.

\section{Remarks}
Using the terminology of this article, in \cite{Goodman} it was stated that for Dirichlet boundary conditions, $\epsilon_r=(-1)^r$. This is a result of a phase change by $-1$ each time the classical path reflects along the boundaries.

In this article we discussed the auxiliary phase that is associated to the boundary point, depending on the boundary condition on the equation imposing on that point being of Dirichlet, Neumann or Robin type. We also proved that the resulting formulae on the propagator matches those arising from the Sch\"odinger equation, with a trivial normalization factor.

There are several generalizations of this result that should be worth investigating:
\begin{enumerate}
\item In this article, we interpret the infinite potential well problem as the Sch\"odinger equation on a line segment with Dirichlet boundary condition and verified that the auxiliary phase is $-1=e^{i\pi}$. It can be shown that for a potential barrier of finite height $h$, the phase factor associated to the reflection should be $e^{-i\theta}$, where $$\theta=\tan^{-1}\left(\frac{k^2-q^2}{2kq}\right)$$
with $k,q$ determined by $h$ and the momentum: $k=\sqrt{\frac{2m}{\hbar^2}E}$ and $q=\sqrt{\frac{2m}{\hbar^2}(h-E)}$.

With the Reflection Principle, the phase factor computed above can then be applied to the problem of a finite potential well. In fact, by doing so, one can obtain its bound states spectra through the sum over classical paths weighted by the corresponding auxiliary phase (which are computed by $\epsilon_r=e^{-ir\theta}$).
\item To study whether the Reflection Principle holds in computing the auxiliary phase on a higher dimensional model when reflecting along the boundary. It would be particularly interesting to know if the geometry (shape or curvature) of the boundary plays a part.
\item Deriving a formula to include the phase factor in this model under other boundary conditions such as the weighted Robin boundary condition (the path integral derivation in cases of the Robin boundary condition with $\psi(0)+\alpha\psi'(0)=\psi(L)+\alpha\psi'(L)=0$ and the mixed condition $\psi(0)=\psi(L)+\alpha\psi'(L)=0$ are also known to us true). In particular, the just-visited phase factor might play a new role in the numerical calculation.
\end{enumerate}

\section*{The Appendix}
In the appendix, we provide heuristics for the auxiliary phase for the Robin boundary condition. Also, we fill in the detailed computation omitted in \S\ref{sec:robin}.

\subsection{Robin boundary condition}
To derive the phase factor for the Robin boundary condition, we consider the Schr\"{o}dinger equation on the half line $\mathbb{R}_{\geq 0}$ and impose the boundary condition $\psi(0)+\alpha\psi'(0)=0$ (the differentiation here is of course just the left derivation).

For an incoming wave with energy $E=\frac{\hbar^2}{2m}k^2$, the wave function should look like
$$\psi(x)=e^{ikx}+Re^{-ikx},$$
where $x\leq 0$. Then
$$e^{ikx}+Re^{-ikx}+\alpha(ike^{ikx}-ikRe^{-ikx})\vert_{x=0}=0.$$
Then $R=e^{i\theta}=-\frac{1-i\alpha k}{1+i\alpha k}$, or equivalently, $\theta=\tan^{-1}(\frac{1-\alpha^2k^2}{2\alpha k})$. One can observe that this phase angle is momentum-dependent, and will change sign when travelling along the opposite direction. In that case, one should consider
$$\psi(x)=e^{-ikx}+Le^{ikx}$$
and $x\geq 0$. In this situation, $L=e^{i\theta}=-\frac{1-i\alpha k}{1+i\alpha k}$ gives mutually inverse angles (the sign here respects the initial direction of the incoming wave). In other words, $L=R^{-1}$. Therefore,
$$\psi(x)=Re^{-ikx}+e^{ikx}$$
is a solution to the Schr\"{o}dinger equation with respect to the boundary condition imposed. This indicates that the phase neutralises for an incoming wave with phase factor $e^{i\theta}$.

Note that when $\alpha$ approaches zero (resp. infinity), the phase factor becomes $-1$ (resp. $1$), which are consistent with the phases implemented in the Dirichlet and Neumann boundary conditions.




\subsection{Handling the wave functions in \S\ref{sec:robin} by cancellation}
For simplicity, we will let $L=1$ in the following computation. From the propagator, we take and simplify the terms
\begin{widetext}
\begin{align*}
&\sum_{n=-\infty}^\infty e^{-ik_nx}\left(e^{ik_ny}+e^{i\theta}e^{-ik_ny}\right)\\
=&\sum_{n=-\infty}^\infty\left(e^{in\pi y}e^{-in\pi x}+e^{i\theta}e^{-in\pi y}e^{-in\pi x}\right)\\
=&\sum_{n=1}^\infty\Big(\big(\cos(n\pi y)\cos(n\pi x)+\sin(n\pi y)\sin(n\pi x)\big)\\
&+\big(\cos\theta\cos(n\pi y)\cos(n\pi x)-\cos\theta\sin(n\pi y)\sin(n\pi x)+\sin\theta\cos(n\pi y)\sin(n\pi x)+\sin\theta\sin(n\pi y)\cos(n\pi x)\big)\Big)\\
=&\sum_{n=1}^\infty\Big(\cos(n\pi y-n\pi x)+\big(\cos\theta\cos(n\pi y+n\pi x)+\sin\theta\sin(n\pi y+n\pi x)\big)\Big)\\
=&\sum_{n=1}^{\infty}\left(\cos(k_ny-k_nx)+\cos(k_ny+k_nx-\theta)\right)\\
=&\sum_{n=1}^\infty 2\cos(k_nx-\frac{\theta}{2})\cos(k_ny-\frac{\theta}{2}).
\end{align*}
\end{widetext}
Note that $\theta$ is an odd function in $k_n$, and therefore in $n$. The result above, when plugged into the propagator, together with the other terms yield (\ref{eq:robin}).

\nocite{*}

\bibliography{refs}{}

\end{document}